\documentclass{aa}
\def\e{et~al.\ }
\def\es{{\rm erg\,s^{-1}} }
\def\gs{{\rm g\,s^{-1}} }

\newcommand{\be}{\begin{equation}}
\newcommand{\ee}{\end{equation}}
\newcommand{\bdm}{\begin{displaymath}}
\newcommand{\edm}{\end{displaymath}}

\begin{document}

   \title{On the accretion luminosity of isolated neutron stars}

   \author{N.R. Ikhsanov\inst{1,2,3}}

 \offprints{N.R.~Ikhsanov\\ \email{ikhsanov@mpifr-bonn.mpg.de}}

   \institute{Max-Planck-Institut f\"ur Radioastronomie, Auf dem H\"ugel 69,
              D-53121 Bonn, Germany
              \and
              Central Astronomical Observatory of the Russian Academy of
              Sciences, Pulkovo 65/1, 196140 St.\,Petersburg, Russia
              \and
              Isaac Newton Institute of Chile, St.Petersburg Branch}
   \date{Received 14 October2002 /  Accepted 6 December 2002}


    \abstract{The accretion process onto a magnetized isolated neutron
star, which captures material from the interstellar medium, is
discussed. The evolutionary track of such a star can be presented
as a sequence of four states: {\it ejector}, {\it supersonic
propeller}, {\it subsonic propeller}, and {\it steady accretor}. I
show that {\it subsonic propeller} $\rightarrow$ {\it accretor}
transition does not occur as long as the magnetic field of the
star is strong enough to control the accretion flow in the stellar
vicinity. During the {\it subsonic propeller} state the accretion
rate onto the stellar surface is limited to the rate of plasma
diffusion into its magnetosphere. The diffusion rate is at least
three orders of magnitude smaller than the maximum possible mass
capture rate by the star. Therefore, the expected accretion
luminosity of magnetized isolated neutron stars is at least three
orders of magnitude smaller than that previously evaluated.
 }

   \maketitle


   \section{Introduction}

The observational appearance of isolated neutron stars (INSs),
which accrete material from the interstellar medium, is a subject
of intensive theoretical investigation (e.g. Shvartsman
\cite{sh70}, Treves \& Colpi \cite{tc91}, Treves et~al.
\cite{ttzc00}, Toropina et~al. \cite{toropina01},  Popov et~al.
\cite{pop00a}). It is presently established that the initially
fast rotating INSs, under certain conditions, are able to switch
their state from {\it ejector} to {\it propeller} within the
Hubble time $t < t_{\rm H}\approx 10^{10}$\,yr, and even to reach
the periods at which the corotational radius,
     \bdm
R_{\rm cor}= (GM_{\rm ns}/\omega^2)^{1/3},
     \edm
exceeds the magnetospheric radius of the star,
  \be
R_{\rm m} \equiv (\mu^{2}/\dot{M}_{\rm c} \sqrt{2GM_{\rm
ns}})^{2/7}.
  \ee
Here $M_{\rm ns}$, $\omega=2\pi/P_{\rm s}$ and $\mu$ are the mass,
the angular velocity, and the dipole magnetic moment of the
neutron star, respectively. $\dot{M}_{\rm c}$ is the mass capture
rate by the star from the interstellar medium, which can be
expressed as
  \be\label{mc}
\dot{M}_{\rm c} \simeq \pi R_{\alpha}^2\ \rho\ V_{\infty}.
  \ee
$V_{\infty}$ is the relative velocity of the neutron star to
the surrounding material, $\rho$ is the density of the
interstellar material, and $R_{\alpha}$ is the accretion radius of
the star,
  \be
R_{\alpha}= 2GM_{\rm ns}/V_{\infty}^2.
  \ee

Limiting $\rho < 10^{-24}\,{\rm g\,cm^{-3}}$ and $V_{\infty} >
V_{\rm s}$, where $V_{\rm s}$ is the thermal velocity in the
interstellar material, Popov \e (\cite{pop00a}) have estimated the
maximum possible rate of mass capture by an isolated neutron star
as
  \be\label{mxc}
\dot{M}_{\rm c} \la \dot{M}_{\rm max} \simeq 10^{12}\ \rho_{-24}\
m^2\ V_6^{-3}\ \gs.
   \ee
Here $m$ is the mass of the star expressed in units of $M_{\sun}$,
$\rho_{-24}=\rho/10^{-24}\,{\rm g\,cm^{-3}}$ and
$V_{6}=V_{\infty}/10^6\,{\rm cm\,s^{-1}}$. Then, assuming that all
material captured by the star is accreted onto its surface, they
have suggested that the accretion luminosity of INSs can be as
high as $\dot{M}_{\rm max}GM_{\rm ns}/R_{\rm ns} \sim 10^{32}
\es$.

The analysis of the validity of this assumption is the subject of
the present paper. I show that under the conditions of interest
the mass accretion rate onto the stellar surface is limited by the
rate of plasma penetration into the stellar magnetosphere, which
is significantly smaller than $\dot{M}_{\rm max}$. Therefore, the
maximum possible accretion luminosity of a magnetized isolated
neutron star proves to be limited to a few\,$\times 10^{28} \es$.
The evolutionary tracks of isolated neutron stars are briefly
reviewed in the next section. The rate of plasma penetration into
the magnetosphere and, correspondingly, the mass accretion rate
onto the surface of isolated neutron stars are evaluated in
Section~3. Here, I also show that the rate of mass accretion onto
the surface of an isolated neutron star remains smaller than
$\dot{M}_{\rm c}$ as long as $R_{\rm m} \gg R_{\rm ns}$. The main
conclusions are summarized in Section~4.

   \section{Evolutionary tracks of INSs}

As was first recognized by Shvartsman (\cite{sh70}), the
evolutionary  track of a rotating magnetized neutron star can be
presented in the form of the following sequence of its states:
{\it ejector} $\rightarrow$ {\it propeller} $\rightarrow$ {\it
accretor}. Within this scheme, the rotational rate of a newly born
fast rotating neutron star decreases, initially by the generation
of the magneto-dipole waves and ejection of relativistic particles
({\it pulsar-like} spin-down), and later by means of the
interaction between its magnetosphere and the surrounding material
({\it propeller} spin-down). The first state transition occurs
when the pressure of the material ejected by the star can no
longer balance the pressure of the surrounding gas, and the
latter, penetrating into the accretion radius of the star,
interacts with the stellar magnetosphere. A detailed analysis of
this state transition with respect to INSs was presented by Popov
et~al. (\cite{pop00b}).

The spin evolution of a spherically accreting strongly magnetized
neutron star in the state of {\it propeller} has been investigated
by Davies, Fabian \& Pringle (\cite{dfp79}) and Davies \& Pringle
(\cite{dp81}). As they shown, two sub-states of the propeller
state can be distinguished: the {\it supersonic} and {\it
subsonic} propeller. In both cases the neutron star is spinning
down due to the interaction between its magnetosphere and the
surrounding gas. As a result of this interaction, the star's
magnetosphere is surrounded by a spherical quasi-static
atmosphere, in which the plasma temperature is of the order of the
free-fall temperature
  \be
T_{\rm ff}(r)=(GM_{\rm ns} m_{\rm p})/(k r).
  \ee
Here $G$, $m_{\rm p}$, and $k$ are the gravitational constant, the
proton mass, and the Boltzmann constant, respectively.

The atmosphere is extended from the magnetospheric boundary up to
the accretion radius of the neutron star. The rotational energy
loss by the neutron star is convected up through the atmosphere by
the turbulent motion and lost through its outer boundary.

The formation of the atmosphere in the first approximation
prevents the surrounding gas from penetrating to within the
accretion radius of the star. As the neutron star moves through
the interstellar medium, the interstellar gas overflows the outer
edge of the atmosphere with the rate $\dot{M}_{\rm c}$ (see
Eq.~\ref{mc}), which is traditionally called {\it the strength of
the stellar wind} and denotes the maximum possible mass capture
rate by the neutron star.

    \subsection{Supersonic propeller}

As long as the angular velocity of the neutron star is large
enough for the corotational radius to be smaller than the
magnetospheric radius, the star is in the centrifugal inhibition
regime (i.e. the centrifugal acceleration at the magnetospheric
boundary, $\omega^2 R_{\rm m}$, dominates the gravitational
acceleration, $GM_{\rm ns}/R_{\rm m}^2$). The centrifugal
inhibition is not effective only within the bases of the
corotational cylinder. However, the accretion of material onto the
stellar surface through these regions does occur only if the the
angle between the magnetic and rotational axes is small enough
(see Ikhsanov \cite{i01c}) and if the magnetic field of the star
is weak enough for the magnetospheric radius to exceed the stellar
radius only by a factor of 2--3~(for discussion see Toropin et~al.
\cite{toropin99} and Romanova et~al. \cite{romanova02}).
Otherwise, the accretion power is significantly smaller than the
spin-down power due to propeller action by the fast rotating star.

Except the bases of the corotational cylinder, the linear velocity
at the boundary of the magnetosphere, which is co-rotating with
the star, in this case is larger than the sound speed in the
atmospheric plasma. That is why this state is usually refereed to
as a {\it supersonic propeller} (see also Ikhsanov \cite{i02}).

    \subsection{Subsonic propeller}

As the star is spinning down, its corotational radius increases
and reaches the magnetospheric radius when $P_{\rm s} = P_{\rm
cd}$, where
 \be\label{pcd}
P_{\rm cd} \simeq 23\ \mu_{30}^{6/7}\ m^{-5/7}\
\dot{M}_{15}^{-3/7}\ {\rm s}.
   \ee
Here $\dot{M}_{15}$ is the strength of the stellar wind expressed
in units of $10^{15}\,{\rm g\,s^{-1}}$, and $\mu_{30}$ is the
dipole magnetic moment of a neutron star expressed in units of
$10^{30}\,{\rm G\,cm^3}$.

Under the condition $P_{\rm s} > P_{\rm cd}$ the centrifugal
barrier is not effective: the atmospheric plasma, penetrating into
the magnetic field of the star, is able to flow along the magnetic
field lines and to accrete onto the stellar surface. However, as
shown by Arons \& Lea (\cite{al76}) and Elsner \& Lamb
(\cite{el76}), the rate of plasma penetration into the
magnetosphere of a spherically accreting strongly magnetized
neutron star can be as high as $\dot{M}_{\rm c}$ only if the
magnetospheric boundary is unstable with respect to interchange
(e.g. Rayleigh-Taylor) instabilities. Otherwise, the rate of
plasma penetration is limited to the diffusion rate, which is a
few orders of magnitude smaller than $\dot{M}_{\rm c}$ (see
Eq.~\ref{dotmdif}). For instability to occur the sign of the
effective gravitational acceleration at the magnetospheric
boundary should be positive:
   \be\label{eff}
g_{\rm eff}\ = \frac{G M_{\rm ns}}{R_{\rm m}^{2}(\theta)}
\cos{\theta} - \frac{V_{\rm T_{\rm i}}^{2}(R_{\rm m})}{R_{\rm
curv}(\theta)} > 0.
   \ee
Here $R_{\rm curv}$ is the curvature radius of the field lines,
$\theta$ is the angle between the radius vector and the outward
normal to the magnetospheric boundary and $V_{\rm T_{\rm
i}}(R_{\rm m})$ is the ion thermal velocity of the accreting
plasma at the boundary. For the case of the equilibrium
magnetospheric shape derived by Arons \& Lea (\cite{al76}), the
condition~(\ref{eff}) can be expressed in terms of the plasma
temperature at the magnetospheric boundary as
   \be\label{cond}
T < T_{\rm cr} \simeq 0.3\ T_{\rm ff}.
  \ee
This indicates that the condition $R_{\rm m} < R_{\rm cor}$ is
necessary, but not sufficient for the effective accretion (with
the rate of $\sim \dot{M}_{\rm c}$) onto the surface of neutron
star to start. In addition, it is required that the cooling of
plasma at the base of the atmosphere is more effective than the
heating.

As shown by Davies \& Pringle (\cite{dp81}), the cooling of the
atmospheric plasma is governed by the bremsstrahlung radiation and
the convective motion. For these processes to dominate the energy
input into the atmosphere due to the propeller action by the star,
the spin period of the star should be $P_{\rm s} \ga P_{\rm br}$,
where $P_{\rm br}$ is a so-called break period, which according to
Ikhsanov (\cite{i01a}) is
   \be\label{pbr}
P_{\rm br} \simeq\ 450\ \mu_{30}^{16/21}\ \dot{M}_{15}^{-5/7}\
m^{-4/21}\ {\rm s}.
      \ee
Under the conditions of interest, the break period significantly
exceeds $P_{\rm cd}$. This means that the state transition {\it
supersonic propeller} $\rightarrow$ {\it steady accretor} may
occur only via an additional intermediate state, which is called
the {\it subsonic propeller}. This term reflects the fact that the
rotation velocity of the magnetosphere during this stage is
smaller than the thermal velocity in the surrounding gas.

If the propeller action were the only source of heating of the
atmospheric plasma, the magnetospheric boundary of the neutron
star would be able to switch its state from {\it subsonic
propeller} to {\it accretor} as its spin period reaches $P_{\rm
br}$. However, as shown below, an additional heating of the
atmospheric plasma occurs due to a radial plasma drift through the
atmosphere. This additional heating mechanism turns out to be not
effective if a star is situated in a relatively strong stellar
wind, but in the opposite case it must be taken into account.

   \section{Diffusion-driven accretion}

Although the interchange instabilities of the magnetospheric
boundary during the subsonic propeller state are suppressed, the
`magnetic gates' are not closed completely: the atmospheric plasma
is able to penetrate into the stellar magnetic field due to
diffusion. The diffusion rate is limited to (see e.g. Ikhsanov
\cite{i01b})
  \be\label{dotmdif}
\dot{M}_{\rm B} \la 10^{11}\,{\rm g\,s^{-1}}\ \zeta_{0.1}^{1/2}\
\mu_{30}^{-1/14} m^{1/7} \left(\frac{\dot{M}_{\rm
c}}{10^{15}\,{\rm g\,s^{-1}}}\right)^{11/14}.
  \ee
Here $\zeta_{0.1}=\zeta/0.1$ is the diffusion efficiency, which is
normalized following the results of experiments on the nuclear
fusion (e.g. Hamasaki \e \cite{ham74}) and the measurements of the
solar wind penetrating the magnetosphere of the Earth (Gosling \e
\cite{gosling91}).

This means that the hot atmosphere surrounding the magnetosphere
of the star in the subsonic propeller state cannot be purely
stationary. For the atmosphere to remain in an equilibrium state,
the amount of material flowing from its base into the
magnetosphere must be compensated for by the same amount of
material coming into the atmosphere through its outer boundary.
Thus the radial drift of plasma through the atmosphere with the
rate $\dot{M}_{\rm B}$ and the velocity
 \be\label{vdr}
V_{\rm dr} = (\dot{M}_{\rm B}/\dot{M}_{\rm c})\ V_{\rm ff}
 \ee
towards the neutron star is expected.

The structure of the atmosphere with the radial plasma drift can
be explain in terms of the quasi-stationary model of Davies \&
Pringle (\cite{dp81}) as long as the characteristic time of the
accretion process,
 \be\label{tdr}
t_{\rm dr} = R/V_{\rm dr},
 \ee
is larger than the characteristic time of turbulent motions
$t_{\rm t} = R/V_{\rm t}$. According to Davies \& Pringle
(\cite{dp81}), the velocity of turbulence motions can be
approximated as $V_{\rm t} \simeq (R_{\rm m}/R)^{1/6} \omega
R_{\rm m}$. Therefore, the condition $t_{\rm t} \ll t_{\rm dr}$
proves be satisfied if the spin period of the star is $P_{\rm s}
\ll P_{\rm qs}$, where
  \be
P_{\rm qs, B} \simeq 1.7 \times 10^5\ \zeta_{0.1}^{-1/2}\
\dot{M}_{15}^{-3/14}\ m^{-6/7}\ \mu_{30}^{13/14}\ {\rm s}.
  \ee
Thus, as long as the spin period of the star is $P_{\rm cd} \la
P_{\rm s} \ll P_{\rm qs, B}$, the magnetosphere of the star is
surrounded by a hot ($T \simeq T_{\rm ff}$) atmosphere in which
the plasma pressure is $p \propto R^{-5/2}$, the sound speed is
$c_{\rm s} \propto R^{-1/2}$ and the number density is $n \propto
R^{-3/2}$ (for discussion see Davies \& Pringle \cite{dp81}).

The radial drift of plasma through the atmosphere towards the
neutron star leads to the release of the accretion (potential)
energy, which is mainly spent in heating the atmospheric plasma.
The heating rate due to this process is $L_{\rm dr}(R)\simeq
\dot{M}_{\rm B} GM_{\rm ns}/R$. This value is small in comparison
with that of the spin-down power (see Davies \& Pringle
\cite{dp81}, Eq.~3.2.4),
   \be\label{lssp}
L_{\rm ssp} = 1.2 \times 10^{36}\ \mu_{30}^2\ m^{-1} P_{\rm
s}^{-3}\ {\rm erg\,s^{-1}},
   \ee
as long as the spin period of the star is $P_{\rm s} \la P_{\rm
acc, B}$, where
  \be\label{paccb}
P_{\rm acc, B} \simeq\ 450\ \zeta_{0.1}^{1/6}\ \mu_{30}^{37/42}\
\dot{M}_{15}^{-5/14}\ m^{-16/21}\ {\rm s}.
  \ee
However, under the condition $P_{\rm s} > P_{\rm acc}$ the heating
of the atmospheric plasma is governed by the accretion power.

The characteristic time of the heating due to accretion process is
$t_{\rm dr}$. On the other hand, the cooling time of the
atmospheric plasma due to the bremsstrahlung emission is
  \be\label{tbr}
t_{\rm cool} \approx t_{\rm br} \simeq 633\,{\rm s}\
\left(\frac{T}{10^9\,{\rm K}}\right)^{1/2}
\left(\frac{n}{10^{13}\,{\rm cm^{-3}}}\right)^{-1},
  \ee
where $n$ is the number density of the atmospheric plasma, which
can be expressed as $n(R)=n(R_{\rm m})(R_{\rm m}/R)^{3/2}$, where
  \be\label{nr}
n(R_{\rm m}) = \frac{\mu^2}{4 \pi R_{\rm m}^6 k T(R_{\rm m})}.
 \ee
Hence, for the temperature of the atmospheric plasma to be smaller
than the critical value, $T_{\rm cr}$, the following condition
should be satisfied
  \be
t_{\rm cool}(R_{\alpha}) < t_{\rm dr}(R_{\alpha}).
  \ee
I require this condition to be satisfied at the outer boundary of
the atmosphere since $t_{\rm c} \propto R$ and $t_{\rm dr} \propto
R^{3/2}$. Combining Eqs.\,(\ref{vdr}), (\ref{tdr}), (\ref{tbr}),
and (\ref{nr}), one finds that the bremsstrahlung cooling
dominates the heating due to accretion power only if the strength
of the stellar wind is $\dot{M}_{\rm c} \ga \dot{M}_0$, where
  \be\label{mzero}
\dot{M}_0 \simeq 3 \times 10^{14}\ \zeta_{0.1}^{7/17}\
\mu_{30}^{-1/17}\ m^{16/17}\ V_{8}^{14/17}\ {\rm g\,s^{-1}}.
  \ee
This means that a magnetized isolated neutron star is able to
switch its state from {\it subsonic propeller} to {\it steady
accretor} only if the mass capture rate by this star from the
interstellar medium is $\dot{M}_{\rm c} \ga \dot{M}_0$. Otherwise,
the corresponding state transition does not occur and the star
remains surrounded by the hot atmosphere. In this case the mass
accretion rate onto the stellar surface is limited to $\dot{M} \la
\dot{M}_{\rm B}$ (see Eq.~\ref{dotmdif}) and, correspondingly, the
accretion luminosity is $L_{\rm x} \la L_{\rm max}$, where
     \be\label{lad}
L_{\rm max} \simeq 10^{30}\ \zeta_{0.1}^{1/2}\ \mu_{30}^{-1/14}\
m^{8/7}\ R_6^{-1}\ \dot{M}_{14}^{11/14}\ {\rm erg\,s^{-1}}.
     \ee
Here $R_6$ is the radius of the neutron star expressed in units of
$10^6$\,cm, and $\dot{M}_{14}=\dot{M}_{\rm c}/10^{14}\,{\rm
g\,s^{-1}}$.

   \section{Discussion}

Application of our findings to the case of INSs leads to the
following conclusions. First, comparing Eqs.~(\ref{mxc}) and
(\ref{mzero}) one finds that under the conditions of interest the
maximum possible strength of the stellar wind of an isolated
neutron star, $\dot{M}_{\rm max}$, is smaller (at least by a
factor of 3) than $\dot{M}_0$. Therefore, the interchange
instabilities of the magnetospheric boundary of these stars are
suppressed and the plasma penetration from the base of the hot
atmosphere into the stellar magnetic field is governed by the Bohm
diffusion. In this case the accretion luminosity of INSs, whose
age is smaller than the characteristic time of the magnetic field
decay, $t_{\rm mfd}$, is limited to
 \be
L_{\rm x}(t<t_{\rm mfd})= \dot{M}_{\rm B} \frac{GM_{\rm
ns}}{R_{\rm ns}} \simeq 3 \times 10^{28}\ \es\ \times
  \ee
  \bdm
\hspace{1cm} \times\ \zeta_{0.1}^{1/2}\ \mu_{30}^{-1/14}\ m^{8/7}\
R_6^{-1}\ \left(\frac{\dot{M}_{\rm c}}{10^{12} {\rm
g\,s^{-1}}}\right)^{11/14}.
  \edm
This radiation is to be emitted from local regions situated at the
magnetic poles of the star. That is why the star could be observed
as a low luminous pulsing X-ray source. The minimum period of this
pulsar is (see Eq.~\ref{pcd})
 \be
P_{\rm min}\ \simeq\ 450\ \mu_{30}^{6/7}\ m^{-5/7}\
\dot{M}_{12}^{-3/7} {\rm s}.
 \ee

Second, as the age of an INS becomes comparable with $t_{\rm
mfd}$, the magnetic field is almost unable to control the
accretion flow in the stellar vicinity and the direct accretion
onto the stellar surface occurs. The accretion luminosity of the
star in this case is $L_{\rm x} = \dot{M}_{\rm c} GM_{\rm
ns}/R_{\rm ns}$, i.e. by a factor of $\dot{M}_{\rm c}/\dot{M}_{\rm
B}$ larger than the accretion luminosity of a neutron star in the
state of the subsonic propeller. This radiation is to be emitted
from the whole surface of the star and, therefore, the pulsations
are not expected to be observed.

Finally, for $\dot{M}_{\rm c} \la 10^{12} {\rm g\,s^{-1}}$, the
spin-down time scale of a neutron star in the subsonic propeller
state is (see Davies \& Pringle \cite{dp81}, Eq.~3.2.5)
 \be
t_{\rm acc, B}\ \ga\  10^9\  \zeta_{0.1}^{-1/2}\  m^{1/7}\
\mu_{30}^{-15/14}\  \dot{M}_{12}^{-3/14}\  P_6\  {\rm yr},
 \ee
where $P_6= P_{\rm acc, B}/10^6$\,s (see Eq.~\ref{paccb}). This
time scale is comparable with the characteristic time scale of the
magnetic field decay (see e.g. Urpin \e \cite{urpin96}). Hence,
the hot atmosphere surrounding the neutron star in the propeller
state can be treated within the quasi-static approximation
suggested by Davies \& Pringle (\cite{dp81}).

 \begin{acknowledgements}
I would like to thank the referee, Dr. Marina Romanova, for useful
comments and suggested improvements. I acknowledge the support of
the Alexander von Humboldt Foundation within the Long-term
Cooperation Program.
\end{acknowledgements}

\end{document}